\documentclass[prl, twocolumn,superscriptaddress,amsmath,amssymb]{revtex4-1}

\usepackage{graphicx}
\usepackage{dcolumn}
\usepackage{bm}
\usepackage{verbatim}
\usepackage{CJK}
\usepackage{amsbsy} 
\usepackage{microtype}
\usepackage{ulem}
\usepackage{tikz}

\begin{document}
\title{Searching for New Spin-Dependent Interactions with SmCo$_5$ Spin Sources and a SERF Comagnetometer}
\author{W. Ji}
\affiliation{Dept. of Phys., Tsinghua University, Beijing, 100084, China}
\author{C. B. Fu}
\email[Corresponding author:]{cbfu@sjtu.edu.cn}
\affiliation{INPAC, Phys. \& Astro. Dept., Shanghai Jiao Tong University, Shanghai, 200240, China}
\author{H. Y. Gao}
\email[Corresponding author:]{gao@duke.edu}
\affiliation{Dept. of Phys., Tsinghua University, Beijing, 100084, China}
\affiliation{Dept. of Phys., Duke University and Triangle Universities Nuclear Laboratory, Durham, NC 27708, U.S.A}
\affiliation{Duke Kunshan University, Kunshan, Jiangsu, 215316, China}

\date{\today}
\begin{abstract}
We propose a novel method to search for possible new macro-scale  spin- and/or velocity-dependent forces (SVDFs)
based on specially designed SmCo$_5$ spin sources and a spin exchange relaxation-free (SERF) comagnetometer. 
A simulation shows that, by covering a SmCo$_5$ permanent magnet with a layer of pure iron,
a high net spin density  source of  about  $1\times 10^{22}$/cm$^3$ could be obtained.
Taking advantages of the high spin density of this iron-shielded SmCo$_5$ and the high sensitivity of the SERF,
the proposed method could set up new limits of greater than 10 orders of magnitude more sensitive than those from previous experiments or proposals in exploring SVDFs in force ranges larger than 1 cm.
\end{abstract}
\maketitle


Searches for anomalous spin- and/or velocity-dependent forces (SVDFs) have drawn considerable attentions in the past few decades.
Various theories beyond the Standard Model have predicted weakly coupled scalar, pseudo-scalar, vector, or axial-vector bosons with light masses\cite{Axion-Review-2010ARNPS, PQ-Axion1977PRL, HILL1988253}.
It is believed that these light bosons may be the answers to many fundamental questions related to, for examples,   
the CP or CPT violation \cite{CPT-PhysRev.136.B1542, CP-Axion-RMP2010},
Lorentz violation \cite{Loong-PRL2016},
and the dark matter\cite{Axion-DM-2001} etc.
Obviously, how to experimentally set limits on coupling constants of such bosons 
or even find them is an important step for human beings to further understand the mother nature. 

The light bosons, if exist, can mediate long-range SVDFs between macroscopic objects\cite{Axion-Review-2010ARNPS}.
Many highly sensitive experimental techniques have been employed to search for these new SVDFs,
for examples, 
the torsion balance \cite{ritter1990experimental, Torsion-Balance2006PRL}, 
the resonance spring \cite{Loong-Nature2003, Loong-PRD2015}, 
the spin exchange relaxation free  (SERF) comagnetometer \cite{Romalis-PRL2009},
and other nuclear magnetic resonance (NMR) based methods \cite{yan2015searching}, etc.

In all these experimental techniques, the spin density of the source is one of the most critical factors.
The force strength mediated by a boson having non-zero mass drops exponentially, 
for example\citep{16Forms2006Dobrescu}, 
\begin{equation}
	V_{2}=\frac{f_{2}\hbar c}{4\pi } (\hat{\boldsymbol\sigma_1}\cdot \hat{\boldsymbol\sigma_2})\left( \frac{1}{r}\right) e^{-r/\lambda},
\end{equation}
where $\hat{\boldsymbol\sigma_{1}},\hat{\boldsymbol\sigma_{2}}$ are the spins of the two particles respectively, $\lambda$ is the interaction range, and $r$ is the distance between the two particles. 
An effective magnetic field 
$B_{eff}=f_{2}\hbar c\, \hat{\boldsymbol\sigma_2} e^{-r/\lambda}/({4\pi r})$ 
can be employed to detect the boson, and increasing the spin density of the source  in the interaction range $\lambda$ can significantly improve the detection sensitivity.
Therefore, various methods have been employed to improve the spin densities
\cite{SEOP-walker1997, Cold-n-2015AB, chu2015search}.

In this letter we propose a new specially designed high spin density material, 
an  iron-shielded SmCo$_5$ permanent magnet (ISSC) together with a SERF comagnetometer to constrain the coupling strengths of the various terms in SVDFs~\cite{16Forms2006Dobrescu}. 
In the following of this letter, we will give an overview of the proposed setup first, 
and  then a short introduction of the SERF comagnetometer.
The structure of the ISSC and its finite element analysis (FEA) simulation are provided.
Then the comparisons between the sensitivities of this proposal and others will be presented.
The limits on the coupling strengths set by this proposal could be improved by as large as more than 10 orders of magnitude compared with those from the previous experiments or proposals.
 
\begin{figure}
\begin{center}
\includegraphics[width=9.cm]{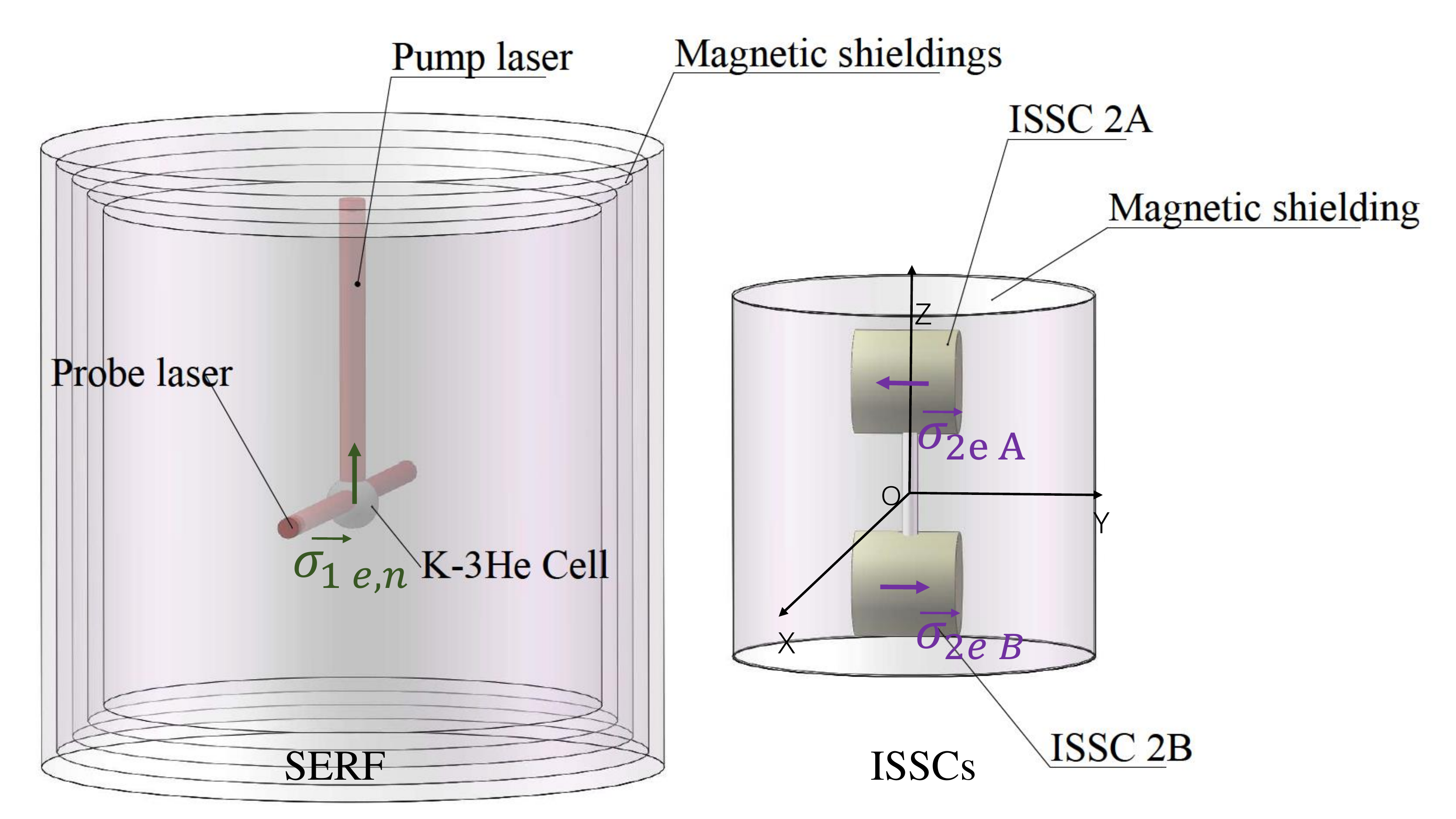}
\caption{
The schematic view of the proposed experiment.  
The neutrons in polarized $^3$He serve as $\vec{\boldsymbol\sigma}_{1n}$;
the electrons in polarized K serve as $\vec{\boldsymbol\sigma}_{1e}$.
Two ISSC spin sources, 2A and 2B,  serve as $\vec{\boldsymbol\sigma}_{2e}^A$ and $\vec{\boldsymbol\sigma}_{2e}^B$. 
By covering ISSCs with an extra layer of $\mu$-metal, the normal magnetic flux leakage from the ISSC can be further reduced while the SVDF signals can still pass through.
By rotating the ISSCs with a given frequency $f_0$ and then locking onto that frequency in the SERF spectra, 
the noises can be reduced, and then the detecting sensitivities can be highly improved. 
Depending on different terms of the SVDF under testing, 
the directions of $\vec{\boldsymbol\sigma}_{2e}^A$  and $\vec{\boldsymbol\sigma}_{2e}^B$  could be put along $x$, $y$, or $z$-direction,  
and the ISSCs can rotate along $x$ or $y$ axis (see text and Tab. \ref{orientation.tab} for details).
}

\label{structure}  
\end{center}
\end{figure}

\paragraph{Experimental Setup--}
The proposed setup is shown in Fig. \ref{structure} schematically. 
The left side is a SERF comagnetometer\cite{allred2002high}.
$^3$He  and K sealed in a glass cell will be polarized and serve as the force probes for the exotic two-body interactions.   
As part of the SERF comagnetometer, several layers of $\mu$-metal cover the K-$^3$He glass cell 
to reduce the possible ambient magnetic fields, and make the system work in the so-called SERF regime. 
The right side in Fig. \ref{structure} is the iron-shielded SmCo$_5$ (ISSC). 
The ISSC can move in different ways,   which will be introduced in details later.
The ISSCs are also covered with $\mu$-metals to further reduce  the magnetic flux leakage from the SmCo$_5$ even after the iron-shielding. 

To get a SERF comagnetometer work, 
the glass cell is normally heated to about 160$^\circ$C to achieve a sufficiently high alkali vapor density. 
The leading order of the alkali atoms' polarization in $x$ direction  is given by\cite{Romalis-PRL2009}:
\begin{equation}\label{pol.eq}
P^{e}_x=\frac{P_z^e  \gamma_e}{R_{tot}} 
\left( 
b_y^n   -  b_y^e
\right),
\end{equation}
where $P^e_i$ is K electrons' polarization along the $i$-axis, 
$b_y^n$ and $b_y^e$ are the magnetic-like field in $y$ direction seen by the $^3$He nucleus and the K electrons respectively, 
$R_{tot}$ is the K electron's relaxation rate, 
and $\gamma_e$ is the gyromagnetic ratio for the K electrons.
Therefore, if the SVDFs exist, and couple to neutrons or electrons, the corresponding effective fields $b_y^n$ or $b_y^e$ can be detected by the SERF comagnetometer.
SERF comagnetometers represent the most sensitive magnetometer today, and sensitivities of about 1 fT/Hz$^{1/2}$ are routinely achievable\cite{Romalis-PRL2009, dang2010ultrahigh}.

\begin{figure}
\begin{center}
	\includegraphics[trim=0 80 0 80,width=5.cm]{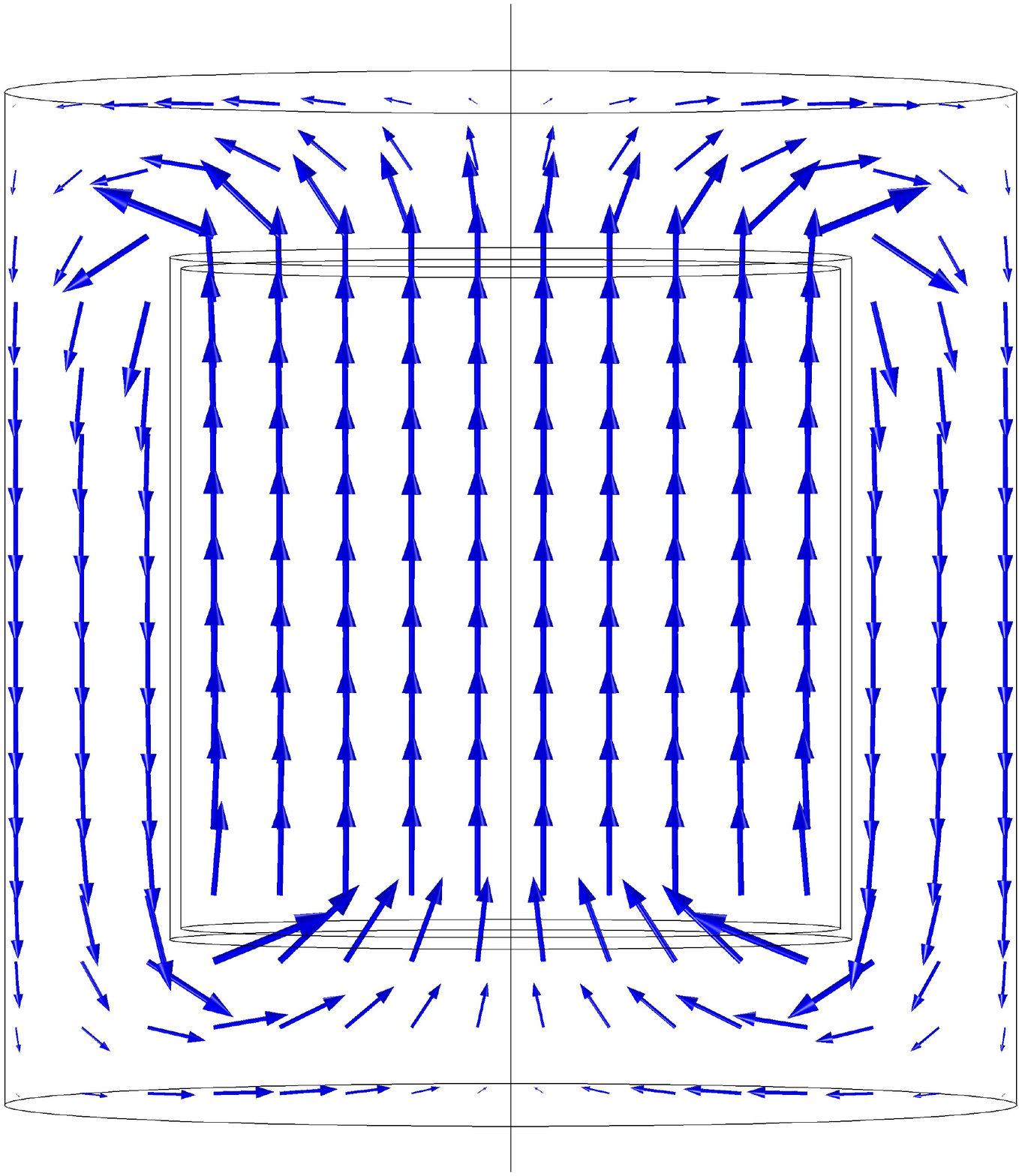}
\caption{A schematic diagram of a spin source and its FEA simulation results.  From inside to outside, there are layers of SmCo$_5$ magnet, air gap, and pure iron.
 The blue arrows represents the magnetic field. In the simulation, the sizes of SmCo$_5$ are set to be $\pi\times 15^2 \times 30$ cm$^3$, and outside pure irons are set to be 7.5 cm thick.
}
\label{spin-source.fig}  
\end{center}
\end{figure}

\paragraph{Iron-shielded SmCo$_5$--}
A higher spin density means a higher sensitivity in the SVDF searching. 
Permanent magnets have high spin densities.
However, a permanent magnet's field can cause large background signals in a SVDF-search experiment if used directly. 
To overcome this, we designed a new structure shown in Fig. \ref{spin-source.fig}. 
At its center, there is a cylindrical SmCo$_5$ magnet.  
Then the SmCo$_5$ cylinder is covered by a layer of pure iron to shield the magnetic field from the SmCo$_5$ core. 

The total electron spin density of ISSC is contributed by two parts: SmCo$_5$ ($ \boldsymbol{n}_{SmCo_5}$) and iron shielding ($ \boldsymbol{n}_{iron}$).
The magnetic moment of the  Sm$^{3+}$ ion is very small at room temperature compared with five cobalt ions, i.e. $-0.04\mu_B$ vs.  $-8.97 \mu_B$\cite{heckel2008preferred, givord1979temperature}. 
It is safe to ignore the magnetization of  Sm in the SmCo$_5$.
Therefore the electron spin density of SmCo$_5$ can be written as\cite{heckel2008preferred},
\begin{equation}\label{eq.ns1}
\displaystyle \boldsymbol{n}_{SmCo}=\frac{f_{Co}(1+R)}{\mu_B}\boldsymbol{\mathrm{M}}_{Co},
\end{equation}
where $\boldsymbol{\mathrm{M}}_{Co}$ is the magnetization of Co, 
$f_{Co}=0.80\pm 0.004$\cite{heckel2008preferred} is the spin contribution of the magnetic moments of Co, 
$R= -0.36$ \cite{heckel2008preferred} is the spin ratio of Sm to Co, 
and $\mu_B$ is the Bohr magneton. 
The minus sign of $R$ means that the Sm spins are in the opposite direction of Co.
For SmCo$_5$ magnetized to 10 kGs, its spin density is about  $ \boldsymbol{n}_{SmCo}=4.5\times10^{22}$/cm$^3$.

The electron spin density of the pure iron can be calculated in the similar way as Eq. \ref{eq.ns1}. 
\begin{equation}\label{eq.ns2}
\boldsymbol{n}_{Fe}=\frac{f_{Fe}}{\mu_B}\boldsymbol{\mathrm{M}}_{Fe},
\end{equation}
where $\boldsymbol{\mathrm{M}}_{Fe}$ is the magnetization of the iron, 
 $\displaystyle{f_{Fe}}=0.957$ is the spin contribution of the magnetic moments in Fe \cite{seavey1958ferromagnetic, frait1971g}. 
The magnetism in pure iron mainly comes from the spin magnetic moment of the $3d$ electrons because the orbital magnetic moment of the $3d$ electrons can be quenched in the inhomogeneous crystalline electric field \cite{bozorth1993ferromagnetism}. 
For pure iron magnetized to 10 kGs, its spin density is about  $ \boldsymbol{n}_{Fe}=8.2\times10^{22}$/cm$^3$.

Even $\mu$-metals may have higher spin densities\cite{shaw2013precise}, we prefer pure iron due to its simple structure which potentially affects the data analysis of the SVDF searching experiments.


\paragraph{FEA Simulation--}
With the FEA method, we simulated the magnetization distribution and then the spin density distributions in the ISSC.
The main optimized input parameters are listed in Tab. \ref{parameters.tab}.
The structure under simulation is shown in Fig. \ref{spin-source.fig}.
The ``net'' electron spin of this ISSC  can be written as 
$\boldsymbol{N}_{net}= \int \boldsymbol{n}_{SmCo}{\rm d} V_{SmCo}+\int \boldsymbol{n}_{Fe} {\rm d} V_{Fe}$,
where $V_{SmCo}$ and $V_{Fe}$ are  the volumes of SmCo$_5$ and Iron respectively.
According to the FEA simulation (shown in Table \ref{parameters.tab}), 
the net electron spin of this structure is $8\times 10^{26}\hbar/2$, 
while at the same time the magnetic field can be cancelled to very close to zero ($<0.5$ Gs at a distance 5 cm away from the pure iron).  

This special feature is mainly due to the fact that Sm's $4f$ electrons and Co's $3d$ electrons in SmCo$_5$ have large orbital magnetic moments, while Fe $3d$ electrons' orbital magnetic moments are quenched.
Therefore, it is possible to reduce the outside magnetic field close to zero, while at the same time keep the total net electron spins non-zero.

\begin{table}
\begin{ruledtabular}
\caption {
The optimized rotating axises and orientations of $\vec{\boldsymbol\sigma}_{2e}^A$ and $\vec{\boldsymbol\sigma}_{2e}^B$ when estimate the sensitivities of the different SVDF terms with the proposed setup.
} 
\label{orientation.tab}
\begin{tabular}{c c c c c c c c c }   

Terms &$V_{2}$&$V_{3}$ & $V_{9+10}$&$V_{11}$&$V_{6+7}$&$V_{14}$&$V_{15}$&$V_{16}$\\
\hline
Rotating axis & $y$ & $y$ & $x$ & $y$ & $x$  & $y$ & $y$ & $y$  \\
$\vec{\boldsymbol\sigma}_{2e}^A$        &$+z$ &$+x$  &$+z$ &$+z$  &$+y$   &$+y$ &$+z$  &$+x$\\
$\vec{\boldsymbol\sigma}_{2e}^B$         &$+z$ &$+x$  &$+z$ &$+z$  &$-y$  &$-y$ &$+z$ &$-x$\\
\end{tabular} 
\end{ruledtabular}
\end{table}

There are some other spin materials which are chosen or proposed for SVDF 
searches, 
for example, Alnico, dysprosium iron garnet (DyIG), and GGG etc\cite{terrano2015short,leslie2014prospects, chu2015search}. 
The Alnico has higher spin densities, but its orbital magnetic moment is too low\cite{heckel2008preferred} to benefit this experiment.
The  garnet-DyIG and GGG crystals have also been used in the SVDF search experiments.
However, fabrication of those crystals are difficult, especially for large-size crystals, which limits their applications.  
Due to the simple structure, stable property, and high spin density, the  ISSC is  an excellent spin material for SVDF searches.
\begin{table}
\begin{ruledtabular}
\caption {Input parameters for the FEA simulation} 
\label{parameters.tab}
\begin{tabular}{c c  c}   
Parameter & value  \\
\hline
The SERF's center to the ISSCs' center& 0.7 m   \\
Distance between the two ISSCs &0.6 m\\
Rotating frequency &5 Hz \\
Soft iron's relative permeability &12000\\
SmCo$_5$ Magnetization  &10 kGs\\
The SERF's sensitivity & 1 fT/Hz$^{1/2}$ \\
Data taking time & 2 weeks \\
The soft Iron's Nucleon density&4.7$\times10^{24}$ cm$^{-3}$\\
The SmCo$_5$'s Nucleon density&5.1$\times10^{24}$ cm$^{-3}$\\
\end{tabular} 
\end{ruledtabular}
\end{table}


\paragraph{Estimations of New Limits for SVDFs--}
By using the ISSC designed above and a SERF comagnetometer, 
the sensitivities of SVDF searches could be estimated.

Mathematically, there are 16 terms of SVDFs\cite{16Forms2006Dobrescu}, 
Here we list the representative 8 terms which are spin-dependent ($V_2$ in Eq.1, $V_3$, $V_{9+10}$,  and $V_{11}$) as well as spin-and-velocity-dependent forces ($V_{6+7}$, $V_{14}$, $V_{15}$,  and $V_{16}$): 
\begin{align}
 \begin{split}
V_{3} =& \frac{f_{3}\hbar^3 }{4\pi m_1m_2 c}
\bigg[
(\hat{\boldsymbol\sigma_1}\cdot\hat{\boldsymbol\sigma_2}) \left( \frac{1}{\lambda r^2} + \frac{1}{r^3}\right)\\
&-(\hat{\boldsymbol\sigma_1}\cdot\hat{\boldsymbol r})(\hat{\boldsymbol\sigma_2}\cdot\hat{\boldsymbol r})
\left( \frac{1}{\lambda^2r }+\frac{3}{\lambda r^2} + \frac{3}{r^3}\right)
\bigg]
e^{-r/\lambda},
 \end{split}\\
V_{\rm 6+7}=&-\frac{ f_{6+7} \hbar^2}{4\pi m_\mu c}\left[(\hat{\boldsymbol\sigma_1} \cdot \mathbf{v})(\hat{\boldsymbol\sigma}_2\cdot\hat{\boldsymbol r})\right]\left(\ \frac{1}{\lambda r}+\frac{1}{r^2} \right) e^{-r/\lambda},
\end{align}
\begin{align}
V_{9+10}=&\frac{ f_{9+10} \hbar^2}{8 \pi m_\mu}(\hat{\boldsymbol\sigma_1} \cdot \hat{\boldsymbol r})
\left(\frac{1}{r\lambda}+\frac{1}{r^2}\right)e^{-r/\lambda} , \\ 
V_{11}=&-\frac{f_{11}\hbar^2}{4\pi m_\mu}
(\hat{\boldsymbol\sigma_1} \times \hat{\boldsymbol\sigma_2})\cdot\hat{\boldsymbol r}]
\left(\frac{1}{r\lambda}+\frac{1}{r^2}\right)
e^{-r/\lambda},\\
V_{14}=&\frac{f_{14}\hbar }{4\pi}\left[(\hat{\boldsymbol\sigma_1}\times\hat{\boldsymbol\sigma_2})\cdot\mathbf{v}\right]\left(\frac{1}{r}\right)e^{-r/\lambda},\\
 \begin{split}
V_{15}=& -\frac{f_{15}\hbar^3 }{8\pi m_1m_2 c^2}
\big\{
(\hat{\boldsymbol\sigma_2}\cdot\hat{\boldsymbol r}) \left[ \hat{\boldsymbol\sigma_1}\cdot(\mathbf{v}\times\hat{\boldsymbol r}) \right]+(\hat{\boldsymbol\sigma_1}\cdot\hat{\boldsymbol r})\\&\left[ \hat{\boldsymbol\sigma_2}\cdot(\mathbf{v}\times\hat{\boldsymbol r})\right]
\big\}
\left( \frac{1}{\lambda^2r }+\frac{3}{\lambda r^2} + \frac{3}{r^3}\right)e^{-r/\lambda},
 \end{split}\\
 \begin{split}
V_{16}=&- \frac{ f_{16}\hbar^2}{8\pi m_\mu c^2}
\big\{ 
(\hat{\boldsymbol\sigma_2}\cdot\mathbf{v})\left[ \hat{\boldsymbol\sigma_1}\cdot(\mathbf{v}\times\hat{\boldsymbol r}) \right]\\
&+(\hat{\boldsymbol\sigma_1}\cdot\mathbf{v})\left[ \hat{\boldsymbol\sigma_2}\cdot(\mathbf{v}\times\hat{\boldsymbol r})\right]
\big\}
\left(\frac{1}{\lambda r}+\frac{1}{r^2}\right)e^{-r/\lambda},
\end{split}
\end{align}
where $f_i$ is the dimensionless coupling constant between particles, $m_1$ and $m_2$ are their respective masses, $m_\mu$ is their reduced mass.

By optimizing the rotational axises of the ISSCs, one can obtain maximum sensitivities for different terms of the SVDFs. 
The ISSCs rotational axises for different SVDF terms are listed in Tab. \ref{orientation.tab}.
The main input parameters, which are conservative, are listed in Tab. \ref{parameters.tab}. 

According to Eq.  \ref{pol.eq}, we take the effective magnetic field for electron as $B^{(e)}_{eff}=b_y^e$, while for neutron, $B^{(n)}_{eff}=b_y^n/0.87$, due to the limited neutron polarization of 87\% in a $^3$He nucleus \cite{friar1990neutron,Romalis-PRL2009}. 

\begin{figure*}[h]
\begin{center}
\includegraphics[width=18.cm,trim=0 50 0 40,height=22.cm]{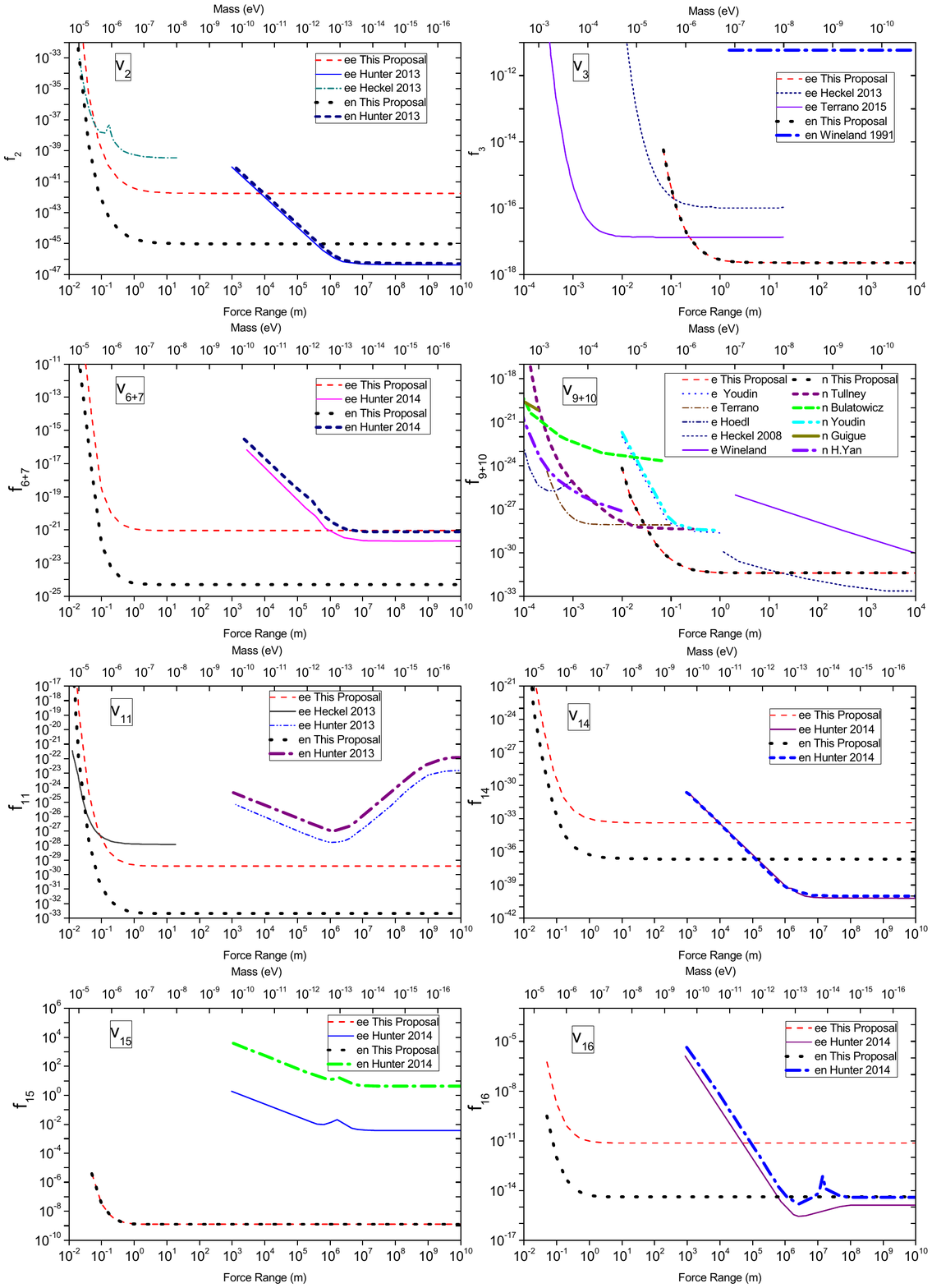}
\caption{
Comparisons of the limits set by this proposal and others in literatures. 
The input parameters, which are conservatively assumed, are shown in Tab. \ref{orientation.tab} \& \ref{parameters.tab}.
The ``ee'' (``en'') labeled here means the coupling between electron and electron (neutron). 
The label ``e'' (``n'') in figure for $f_{9+10}$ means coupling between unpolarized mass and electron (neutron).
The references for different terms of the SVDFs are:
$V_2$ from Ref. \cite{heckel2013limits,hunter2013using,kotler2015constraints}, 
$V_3$ from Ref. \cite{kotler2015constraints, terrano2015short,wineland1991search,
heckel2013limits},
$V_{6+7},V_{14},V_{15}$and$ V_{16}$ from Ref~\cite{hunter2014using},
$V_{9+10}$ are from Ref~\cite{youdin1996limits,terrano2015short,ni1999search,hammond2007new,
hoedl2011improved,tullney2013constraints,serebrov2010search,
petukhov2010polarized,bulatowicz2013laboratory,chu2013laboratory,
heckel2008preferred,yan2014probing},
and $V_{11}$ from Ref~\cite{hunter2013using,heckel2013limits}.
}
 \label{results.fig}  
\end{center}
\end{figure*}

The estimated results are shown in Fig. \ref{results.fig}. 
In $\lambda >0.1$ m, even with conservative input parameters, 
the proposed method could highly improve the sensitivities of these types of the SVDF searches. 
For example, the limits of $f_{15}^{(en)}$, and $f_{16}^{(en)}$  can be improved by over 10 orders of magnitude  in $\lambda<1000$ m compared with the current best limits\cite{hunter2014using}. 
For the constrains of the possible new ``dipole-dipole interaction'' $f_{3}^{(en)}$ and $f_{6+7}^{(en)}$, this proposal could be more than 7 orders of magnitude better than the current records\cite{wineland1991search,hunter2014using} at around $\lambda=1000$ m.
For the limits of $f_{9+10}$ and $f_{2}^{(en)}$, this proposal could be more than over 3 orders of magnitude better than other proposals\cite{leslie2014prospects,chu2015search}.
For other terms of the SVDFs for electron-electron (ee) and electron-nucleon (en) couplings, the proposed method can also be several orders of magnitude better than other corresponding best limits up to date.


\paragraph{Summary--}
The experimental searches for new macro scale SVDFs are important for testing theories beyond the Standard Model. 
High spin density and easily handling materials are critical for SVDF-search experiments.
We propose the ISSC structure, i.e. a SmCo$_5$ permanent magnet covered with pure iron,  for the SVDF studies. 
In this new structure, the magnetic field could be highly reduced, 
while at the same time a large amount of net electron spin polarization can be achieved.
By using this ISSC structure, together with the highly sensitive SERF comagnetometer, 
the sensitivities for detecting different terms of SVDFs are discussed.
This new approach has sensitivities as large as 10 orders of magnitude higher compared with those from previous experiments or proposals, which makes it a promising method in new experiments searching for spin-dependent interactions. 

\paragraph{Acknowledgement--}
This work is supported by 
the National Nature Science Foundation of China (Grant No. 11375114),
and the US Department of Energy under contract number DE-FG02-03ER41231. 
We thank Y. Chen for useful discussion on SERF magnetometers.
One of us (CBF) thanks Shanghai Municipal Science and Technology
Commission for the supports (under grant No. 11DZ2260700)).

\bibliographystyle{apsrev4}
\bibliography{5thForce}

\end{document}